\documentclass{PoS}
\title{Effective Field Theory for Long Strings}
\ShortTitle{Effective Field Theory for Long Strings}
\author{{M. Baker}\\
     Department of Physics, University of Washington, Seattle\\Box 351560, Seattle WA 98195, USA\\
     E-mail: \email{mbaker4@u.washington.edu}}
    
\abstract{In previous work we used magnetic SU(N) gauge theory with adjoint representation Higgs scalars  to describe the long distance quark-antiquark interaction in pure Yang-Mills theory, and later to obtain an effective
string theory. The empirically determined parameters of the non-Abelian
effective theory yielded $Z_N$ flux tubes resembling those of the Abelian 
Higgs model with Landau-Ginzburg parameter  equal to $ 1/\sqrt{2}$, 
corresponding to a superconductor on the border between type I and type II.
However, the physical significance of the differences between the Abelian 
and the $Z_N$ vortices was not elucidated and no principle was found to 
fix the value of the 'Landau-Ginzburg parameter'  $\kappa$ of the non-Abelian 
theory determining the structure of the $Z_N$ vortices. 

Here we reexamine this point of view.  We propose a consistency condition 
on  $Z_N$ vortices underlying a confining
string.  This fixes the value of $\kappa$. The transverse distribution of pressure  $p(r)$ in the resulting
$Z_N$ flux tubes
provides a physical picture of these  vortices which
differs essentially from that of the vortices of the Abelian Higgs model. We speculate that this  general picture is valid independent of the  details of the effective magnetic gauge theory from which it was obtained. 
Long wavelength fluctuations of the axis of the $Z_N$ vortices
lead from an effective field
theory to an effective string theory with the Nambu-Goto action. This effective string theory
depends on a single 
parameter, the string tension $\sigma$.  In contrast, the effective field 
theory has a second parameter, the intrinsic width $1/M$ of the
 flux tube,
and is applicable at intermediate distances in a 
range between $0.2\,fm$ and $1\,fm$, where the contribution of the 
intrinsic width increases the flux tube width over that predicted by 
effective string theory.  
}
       
\FullConference{Xth Quark Confinement and the Hadron Spectrum\\
              8--12 October 2012\\
              TUM Campus Garching, Munich, Germany}

\begin{document}
\sf
\maketitle

\sf

\section {Introduction}

The principal goal of this talk is to reexamine magnetic SU(N) gauge theory
 which we have used \cite{BBZ:1990}
as an effective field theory of the long distance quark-antiquark 
interaction, and to elucidate the properties
of the $Z_N$ flux tubes found in the theory,

In Sections 2 to 4 we write down the Lagrangian of the effective $SU(N)$
gauge theory. We obtain a relation, applicable for any configuration of the Higgs fields,
 between the string tension $\sigma$ and the 
transverse distribution of pressure $p(r)$ in the resulting $Z_N$ flux tubes. 
In section 5, using this relation, we impose a constraint
on $p(r)$, $\int_0^\infty r p(r) dr =0$,
which we speculate is a necessary condition
for a flux tube to behave as a string.  

In sections 6 and 7 we consider $SU(3)$, where we have found
explicit classical $Z_3$ flux tube solutions, and we compare
these solutions to those found in the Abelian Higgs model.
We plot the pressure distributions $p(r)$ in the $Z_3$ flux tube,
and describe the physical picture that emerges.
The pressure is positive near the axis and at larger distances it is negative. It is natural to associate the boundary between the outside and inside of the string with the point at which the pressure vanishes.

 In sections 8 and 9 we show that long wavelength fluctuations of the flux tube axis  lead  
from an effective field theory to an effective string
theory 
with a single parameter, $\sigma$. The contribution of these fluctuations to the flux tube width \cite{Gliozzi} fixes the value of the short distance cutoff $1/\Lambda$ of the effective field theory at a value less than the intrinsic width $1/M$. Thus the theory can  resolve distance scales on the order of $1/M$.
Finally, we examine the impact of string fluctuations on the domain of applicability of the  effective field theory.


\section {Effective Field Theory }

The Lagrangian ${\cal{L}}_{eff}$ couples magnetic $SU(N)$ gauge potentials, $C_\mu$ to three adjoint representation scalar 
fields $\phi_i$.  The gauge coupling constant  is $g_m$ .
\begin {equation}
{\cal{L}}_{eff} (C_\mu, \phi_i) = 2 tr (-\frac{1}{4} G^{\mu \nu} G_{\mu \nu} + \frac{1}{2} ({\cal{D}}_\mu \phi_i)^2) - V(\phi_i),
\label{Leff}
\end{equation}
\begin {equation}
G_{\mu \nu} = \partial_\mu C_\nu - \partial_\nu C_\mu - i g_m [C_\mu,\,C_\nu] ,\,\,\,\,{\cal{D}}_\mu \phi_i = \partial_\mu \phi_i - i g_m [C_\mu,\,\phi_i].
\label{Gmunu}
\end{equation}
The components of the field tensor $G^{\mu \nu}$ define color electric and magnetic fields $\vec{E}$ and $\vec{B}$.
\begin {equation}
E^k = \frac{1}{2} \epsilon_{klmn} G^{lm},\, B^k = G^{k0}.
\label{EB}
\end{equation}
The Higgs potential $V(\phi_i)$ is generated from one loop graphs 
of SU(N) gauge theory:\cite{BBZ:1990}
\begin{equation}
V(\phi_i) = \frac {\mu^2 N}{4}  \sum_i 2~ 2 tr \phi_i^2  +
\frac {4 N \lambda}{3} \left ( tr( \sum_{i\,j}  \phi_i^2 \phi_j^2) + \frac{1}{N}(tr ( \sum_i  \phi_i^2) )^2 + \frac{2}{N} \sum_{i\,j}( tr \phi_i \phi_j )^2 \right),
\label{Vphii}
\end{equation}
where the parameter $\mu^2$ has dimensions of mass squared and $\lambda$ is dimensionless. The SU(N) gauge symmetry of ${\cal{L}}_{eff}$ reflects that of the original SU(N) Yang-Mills theory.

 In the confining vacuum  the magnetic SU(N)
/$Z_N$ gauge symmetry is  completely broken by a  Higgs condensate 
$<\phi_i> = \phi_{i0}
 = \phi_0 J_i,$
where the three matrices $J_i$ are the generators of N-dimensional irreducible representation
 of the three-dimensional  rotation group. The  Higgs potential has  an absolute minimum at $\phi_i = \phi_{i0}$: 
$\phi_0^2 = - \frac{9 \mu^2}{8 (N^2 -1 ) \lambda}$.

\section {$Z_N$ Electric Flux Tubes}

At large distances $r$  from  the flux tube axis
 $\phi_i$ and $C_\mu$ are a gauge transformation  $\Omega(\theta) $ of the vacuum $\phi_i = \phi_{i0},\,\, C_\mu =0$, which we can choose  to be Abelian; $\,\,\,\Omega(\theta) = exp(i \theta Y).$
\begin{equation}
\phi_i  \rightarrow \Omega^{-1} (\theta)  \phi_{i 0} \Omega (\theta),\,\,C_\mu \rightarrow \frac{i}{g_m} \Omega^{-1} (\theta) \partial_\mu \Omega(\theta). 
\label{phii}
\end{equation}
 The requirement that $\phi_i$ be  single valued $\rightarrow \,\,\,\,exp(i\,2\,\pi Y)$ is an element of $Z_N$. 
 
As $ r \rightarrow \infty$
\begin {equation}
\vec{C} \rightarrow \frac{1}{g_m,r} \hat{e}_{\theta} Y, \,\,\,\,\,\,\,\,\,exp( i g_m \oint \vec{C} \cdot d\vec{l}) \rightarrow exp (2 \pi i  \frac{k}{N}),\,\,k=1,2, N-1.
\label{vecC}
\end{equation}
Assuming the gauge potential $\vec{C} = C(r) \hat{e}_\theta Y$ everywhere
implies that the electric field  
\begin{equation} 
\vec{E} = - \nabla \times \vec{C}(\vec{x})\, Y = -\frac{1}{r} \,\frac{d (r C(r))}{dr}\hat{e}_z \,\,Y. 
\label{vecE}
\end{equation}
The finiteness of the flux tube energy $\rightarrow \phi_i =0$ on the flux tube axis.

 \section { Relation Between String Tension and Stress Tensor in SU(N) Flux Tubes }
 
  Using the Abelian ansatz (\ref{vecE}) and the resulting  classical static equation
\begin {equation}
\nabla \times \vec{E}\,\, =
-\vec{j}\,\, =  i g_m [\phi_i,\vec{\cal{D}} \phi_i]
\label{j}
\end{equation}
to evaluate ${\cal{L}}_{eff}$ gives the following general  relation between the string tension $\sigma$,  the stress tensor component $T_{\theta \theta}$, and  $\vec{E} (r=0)$, the color electric field on the axis of the flux tube: 
 \begin{equation}
 \int_0^\infty 2 \pi r  \frac{T_{\theta \theta} (r)}{r^2} dr = \,-\,2\, tr ( \frac {2 \pi}{g_m} Y\, \hat{e} \cdot \vec{E} (r=0) )  - \sigma. 
 \label{sigmaW}
 \end{equation}
valid for  any configuration of the Higgs fields $\phi_i$.
The quantity $ -2\, tr ( \frac {2 \pi}{g_m} Y\,\hat{e_z}\cdot\vec{ E} (r=0) ) R = W$, 
the work necessary to separate a $q\,\bar{q}$ pair lying on
the z-axis by a distance $R$.
  
If $T_{\theta \theta} > 0 $  the gauge repulsion exceeds the Higgs attraction produced by the circulating magnetic currents  generated by the Higgs condensate, and  (\ref{sigmaW}) implies that $ W >  \sigma\,R$. That is,
when there is  net repulsion, the work $W$ needed to separate the $q\, \bar{q}$ pair a distance $R$ in the fixed  final vortex field $\vec{E} (r=0)$ is greater than
 $ \sigma R$, which itself is equal to the work done in 
a field $\vec E $ that  is being built up as the  $q\,\bar{q}$ pair 
is separated.

If there is compensation between the net attractive and repulsive contributions to the pressure  $p(r) =T_{\theta \theta}/r^2$ averaged over the width of the flux tube; that is, if

\begin{equation}
\int_0^\infty 2 \pi r \frac{T_{\theta \theta}}{r^2} dr  = \int_0^\infty 2 \pi r p(r) dr =0,
\label{balance}
\end{equation}
then the string tension $\sigma = W/R$, determined by the field $\vec{E} (r=0)$ on the flux tube axis.

\section{Speculation on Effective Field Theories Describing Long Strings}

Consider now  a flux tube connecting  a $q\,\bar{q}$  pair located at 
 $z = \pm R/2$ having energy $V_0 (R)$, the heavy quark potential.
  The force acting on the quarks is determined by the color field  
at the positions of the quarks and is  equal to $d V_0 /dR$.
 If the long distance potential $V_0 (R) = \sigma R$ persists to 
distances $R $, then this field is fixed by the string
tension $\sigma$. 

If condition (\ref{balance}) is met, the  field $\vec E$
on the $z$ axis  near the midpoint of the flux tube is  
also fixed by the value of $\sigma$.  In this situation, it is consistent
to assume the field  has the same value, proportional to $\sigma$,
at all points on the $z $ axis between the $q\bar q$ pair   
; that is, the flux tube behaves like a string, consistent with the assumption that  the long distance $q\,\bar{q}$ interaction persists to short distances.
This argument fails if condition (\ref{balance}) is not satisfied.

We assume that (\ref{balance}) must be satisfied
for any effective field theory describing the confining string in SU(N)
Yang-Mills theory, and we impose this condition 
to constrain the parameters in ${\cal{L}}_{eff}$.

\section {Classical Static  SU(3) Flux Tube Solutions }
For SU(3) we have found explicit classical static solutions \cite{BBZ:1990} with
\begin {eqnarray}
\label{newphi}
J_x & =& \lambda_7,\,\,\,J_y = -\lambda_5,\,\,\,J_z =  \lambda_2,\,\,\,Y = \frac{\lambda_8}{\sqrt{3}},\\
\nonumber  
\phi_1 & = & \phi_1(\vec{ x}) \frac{(\lambda_7 - i \lambda_6)}{2} + \phi_1^* (\vec{x}) \frac{(\lambda_7 + i \lambda_6)}{2},\\  \nonumber
\phi_2 & = &\phi_2(\vec{x}) \frac{(-\lambda_5 - i \lambda_4)}{2} + \phi_2^* (\vec{x}) \frac{(-\lambda_5 + i \lambda_4)}{2},\\  \nonumber
\phi_3 & = &\phi_3 (\vec{x}) \lambda_2, \\
\nonumber
 \vec{C} &  = &C(r) \hat{e}_\theta Y,\,\,\,\phi_1 (\vec{x}) = \phi(r) exp (-i \theta),\,\,\, \phi_2 (\vec{x}) = \phi(r) exp (i \theta),\,\,\,\phi_3 (\vec{x}) = \phi_3 (r).
\end{eqnarray}
The commutation relations
\begin{equation}
[Y, \lambda_7 - i \lambda_6] = \lambda_7 - i \lambda_6, \,\,\, [ Y, -\lambda_5 - i \lambda_4] = - (-\lambda_5 - i \lambda_4),\,\, [Y, \lambda_2] =0
\end{equation}
show that the Higgs fields $\phi_1\, \phi_2,\,\phi_3$ carry $Y$ charge $-1,\, 1,\,0$ respectively, and that the ansatz (\ref{newphi}) is consistent with equation (\ref{j}).

We rescale the fields choosing the flux tube radius $\frac{1}{M}$
as the scale of length, making the replacement
$ r \rightarrow r/M,\,\, C \rightarrow \frac{M C}{g_m}, \,\, \phi \rightarrow \phi_0 \phi, \,\, \phi_3 \rightarrow \phi_o \phi_3, $
with $M = \sqrt{6} g_m \phi_0$. Inserting  (\ref{newphi})
 into the effective Lagrangian (\ref{Leff})
yields  the energy density $T_{00}$ and the stress tensor component $T_{\theta \theta}$:

\begin {eqnarray}
\label {Tthetathetanew}
T_{00} & = & \frac{4}{3} \frac{M^2}{g_m^2}  \left [ \frac{1}{2} (\frac{1}{r} \frac {d(r C)}{dr} )^2 + \frac{1}{2} (C - \frac{1}{ r})^2 \phi^2 + \frac{1}{2} (\frac{d\phi}{dr})^2 + \frac{1}{4}  (\frac{d \phi_3}{dr})^2  + V (\phi, \phi_3) \right ],\\
\frac{T_{\theta \theta}}{r^2} &= &\frac{4}{3} \frac {M^4}{g_m^2} \left [ \frac{1}{2} (\frac{1}{r} \frac{d (r C)}{dr} )^2 + \frac{1}{2} (C - \frac{1}{r})^2 \phi^2 - \frac{1}{2} (\frac{d \phi}{dr})^2 - \frac{1}{4} (\frac {d \phi_3}{d r} )^2 - V (\phi, \phi_3) \right ]. 
 \end{eqnarray}
 where
 \begin{equation}
V(\phi, \phi_3)
= \kappa^2 \left (   \frac{ (\phi^2 -1)^2}{4}  +9 \frac{ (\phi_3^2 -1 )^2}{100} - 7  \frac { (\phi_3^2 -1) (1 - \phi^2)}{50} \right ),\,\,\,\,
 \kappa^2 \equiv  \frac{25}{9} \frac {\lambda}{g_m^2}.
 \label{Wnew}
 \end{equation}

Separating the gauge contribution  and the Higgs contribution
to  $T_{00}$ and $T_{\theta \theta}$ 
 gives
 
\begin {eqnarray}
   \int_0^\infty 2 \pi r T_{0 0} dr  = \sigma  =\frac{4}{3}\frac{M^2} {g_m^2} \sigma(\kappa) & =& \frac{4}{3} \frac{M^2}{g_m^2} (\sigma_g (\kappa) + \sigma_h (\kappa)), \\
    \int_0^\infty 2 \pi r \frac{T_{\theta \theta}}{r^2} dr &=& \frac{4}{3} \frac {M^2}{g_m^2} (\sigma_g (\kappa) - \sigma_h (\kappa)),
  \label{sigmadivided}
  \end{eqnarray}
  where
  \begin{equation}
  \sigma_g (\kappa) \equiv
\int_0^\infty \,\,2\pi  r\,dr\,   \left ( \frac{1}{2} (\frac{1}{r} \frac {d(r C)}{dr} )^2 + \frac{1}{2} (C - \frac{1}{ r})^2 \phi^2 \right),
\label{sigmag}
\end{equation}
and
\begin{equation}
\sigma_h (\kappa) \equiv \int_0^\infty \,\,2 \pi r\,dr\,  \left ( \frac{1}{2} (\frac{d\phi}{dr})^2 + \frac{1}{4}  (\frac{d \phi_3}{dr})^2  + V (\phi, \phi_3) \right).
\label{sigmah}
\end{equation}

\section {Results for SU(3) String Tension and Stress Tensor}

\begin{figure}[htbp]
\begin{center}
\includegraphics[width=4in]{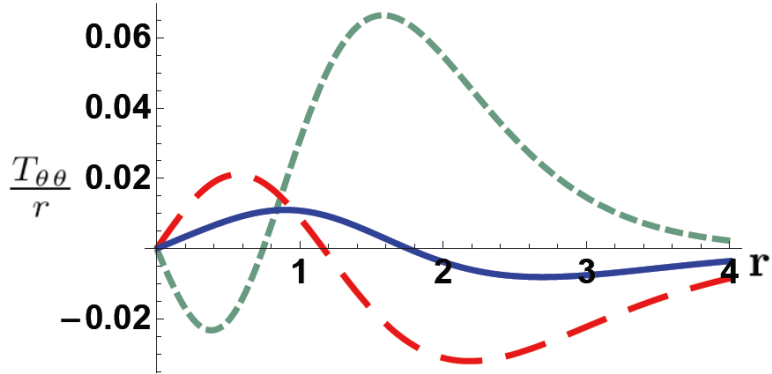}
\caption{ $T_{\theta\theta}/r$ vs $r$.  Red, long dashed, $\kappa^2 = 0.5$;  ~~ blue, thick, $\kappa^2= 0.59$; ~~  green, short dashed, $\kappa^2 = 0.8$. }
\label{3Ttth}
\end{center}
\end{figure}

Note that if $\phi_3(r)$ has its vacuum value $\phi_3 =1,\,\, V(\phi, \phi_3)$  reduces to the Higgs potential of the Abelian Higgs model with Landau Ginzburg parameter $\kappa$.  Furthermore, numerical solution of the classical equations shows that  $\phi < 1$ and $\phi_3  >1 $ everywhere;  hence the term coupling  $\phi$ and $\phi_3$  in (\ref{Wnew})  is attractive.  This  additional attraction   in $V(\phi, \phi_3)$ reduces the energy of the $Z_3$ vortex  below that of the Abrikosov-Nielsen-Olesen vortex 
of the Abelian Higgs model.  The ANO vortex can then be
viewed as an unstable configuration of the non-Abelian theory that subsequently decays to the stationary classical solution  $\phi_3 (r). $

Condition (\ref{balance}) along with (\ref{sigmadivided}) yield $\sigma_g (\kappa) =\sigma_h (\kappa)$, which determines the physical value of $\kappa$; $\kappa^2 \approx 0.6$.  The string tension   $\sigma(\kappa^2 \approx 0.6) \approx 3.1$,  approximately equal to its value  in the Abelian Higgs model at $\kappa^2 =1/2 $.  Figure 1 shows $T_{\theta \theta}/r$ evaluated at the classical solution as a function of $r$  for three values of $\kappa^2$.  
For $\kappa^2 \approx 0.6$, where condition (\ref{balance}) is satisfied,
$T_{\theta \theta } =0 $  at $r  \equiv r* \sim 1.7/M $; there is repulsion at $r \le r*$ and attraction at  $r > r*$.
 It is natural to identify $r*$ as a boundary, separating the inside  of the flux tube  from its exterior.

In contrast, in the Abelian Higgs model 
$\kappa =\frac{1}{\sqrt{2}}$ is a BPS state \cite{bogomolny}, and condition (\ref{balance}) is satisfied exactly because 
the components $T_{\theta\theta}$ and
$T_{rr}$ of the stress tensor 
vanish for all $r$ \cite{deVega},
and thus the profile  of $T_{\theta \theta} (r)$ does not reveal  the boundary of the flux tube.

We speculate that the difference between the properties of the stress tensor inside (positive net pressure) and outside the flux tube (negative net pressure) is a fundamental physical property of flux tubes  giving rise to a confining string. 
The difference between the Abelian and non-Abelian theories is caused by the additional attractive interaction between the scalar particles, which breaks the supersymmetry \cite{fayet} giving rise to the BPS Abelian Higgs vortex,  and stabilizes the non- Abelian flux tube.
Indeed,  as we have seen, the additional attraction in the Higgs potential of the non-Abelian theory
 is approximately balanced  at $\kappa^2 \approx 0.6$ by the additional repulsion associated with the fact that $\kappa^2 > 1/2$.

\section{ From Effective Field Theory  to Effective String Theory}

The Higgs fields $\phi$ vanish on the axis $L$ of the static flux tube.
Long wavelength fluctuations of  the axis $L$ of a flux tube connecting a quark-antiquark pair sweep out  a space time surface $\tilde{x}^\mu (\zeta)$ on which $\phi$ vanishes.
The Wilson loop $W(\Gamma)$ of Yang-Mills theory is the path integral over all field configurations for which the Higgs fields  vanish on some surface $\tilde{x}^\mu (\zeta)$ whose boundary is  the loop $\Gamma$. 

\begin{equation}
W(\Gamma) = \int{\cal{D}} C_\mu{\cal{D}} \phi \,exp(i S(C_\mu, \phi),\, \,\,\,\,\, S(C_\mu, \phi) = \int dx{ \cal{L}}_{eff} (C_\mu, \phi).
\label{W}
\end{equation}

We transform $W(\Gamma)$ to a path integral over the vortex sheets $\tilde{x}^\mu (\zeta)$ in two stages:
\begin {enumerate}
\item 
 We fix the location $\tilde{x}^\mu (\zeta)$ of the vortex and integrate  over field configurations $C_\mu (x),\,\, \phi (x)$ for which $\phi (x)|_{x= \tilde{x} (\zeta)} =0.$ The integration (\ref{W}) over these configurations  $\rightarrow \,S_{eff} (\tilde{x})$, the action of the effective string theory.

\item We then integrate over all surfaces $\tilde{x}^\mu (\zeta)$. This integration puts  $W(\Gamma)$ into the form of  a partition function of an effective string theory: \cite{Baker+Steinke}

\end{enumerate}
\begin {equation}
W(\Gamma) = \int {\cal{D}}\tilde{x}^\mu\,.\,.\,.exp[iS_{eff} (\tilde{x}^\mu)].
\label{Wagain}
\end{equation}
The path integral  (\ref{Wagain}) goes over the two transverse fluctuations of the world sheet $\tilde{x}^\mu (\zeta)$.

The field modes contributing to $S_{eff} [\tilde{x}^\mu]$ have masses $> M$. 
 Fluctuations of wavelength $ > 1/M$ are string fluctuations accounted for by (\ref{Wagain}).
We can then replace integrations (\ref{W}) over field configurations $C_\mu,\,\phi$ by the  classical configuration minimizing $S (C_\mu,\,\phi)$ for fixed position $x^\mu (\zeta)$ of the vortex.

$$exp(iS_{eff} (\tilde{x}^\mu(\zeta))) \approx exp (i S^{class} (\tilde{x}^\mu(\zeta))).$$

When condition (\ref{balance}) is satisfied,  the linear potential persists when a straight flux tube is shortened. Likewise,  bending the flux tube slightly gives a change in energy proportional to the change $\Delta R$ in length: $ \Delta E = \sigma\,\, \Delta R$.   The action of the 
effective of the effective field theory becomes the Nambu-Goto action  
proportional to the area of the vortex sheet.

\begin {equation}
S_{eff} (\tilde{x}^\mu) = \sigma \int d^2 \xi  \sqrt{-g(\xi)} \equiv S_{NG} (\tilde{x}^\mu).
\label {newaction}
\end{equation}

\section {Heavy Quark Potentials and Flux Tube Shape}

To obtain  the heavy quark potential $V_0 (R)$ and transverse energy
profiles between static quarks separated by distance $R$ we couple the
vector potential $\vec{C}$ to  a Dirac string, writing
\begin {equation}
\vec{E} = - \nabla \times \vec{C}  - \frac{2 \pi}{g_m} \delta (x) \delta (y) (\theta (z+R/2) - \theta (z-R/2))\hat{e}_z Y
\label{oldE}
\end{equation}
in the Lagrangian  ${\cal{L}}_{eff}$, and solving the resulting static equations  \cite{BBZ:1991}.  We
compared the results  \cite{BBZ:1997} with lattice data for heavy quark potentials  \cite{bali}, and  found that $g_m \approx 3.91$; i. e., $M \approx 1.9 \sqrt{\sigma}$.  
Furthermore, these calculations were consistent with SU(2) lattice simulations \cite{green} for  transverse energy profiles for a range of  interquark spacings $0.25/\sqrt{\sigma} \le R <  2/\sqrt{\sigma}$.

 The above calculations did not explicitly include the contribution of string fluctuations. However, string fluctuations renormalize the intrinsic width and therefore they are to some extent accounted for 
in the empirically determined value of $M$. For  distances larger than  $ \sim \,1/\sqrt{\sigma}$, string fluctuations become dominant, leading to the  logarithmic increase of the mean square width of the flux tube at its midpoint \cite{Munster}; 

 \begin{equation}
 w^2 (R/2) = \frac {d-2}{2 \pi \sigma}\, log \,\frac{R}{r_0}.
 \label{logincrease}
 \end{equation}

Recent  lattice simulations of $(2+1)\,d \,SU(2)$ Yang-Mills theory \cite{Gliozzi}  extending to distances $R= 36/\sqrt{\sigma}$ gave excellent agreement with the predictions  of effective string theory for distances $R > 1.5/\sqrt{\sigma}$, and determined the value of  $r_0 =  0.364/\sqrt{\sigma}$. (Interpreting
$ 1/r_0= \Lambda  $ as  the  cutoff of the effective field theory gives  $ \Lambda \approx 2.75 \sqrt{\sigma} \approx 1.41 M $.)  However, for distances $1.5/\sqrt{\sigma } > R > 0.2/\sqrt{\sigma}$ the lattice simulations of $w^2(R/2)$  lie above the prediction (\ref{logincrease}), indicating that the intrinsic width of the flux tube must be taken into account at these $q \,\bar{q}$ separations.   It is in this intermediate range, shown schematically  in Figure 2,  that we can test the physical picture of the confining string given by the effective field theory.
 \begin{figure}[htbp]
\begin{center}
\includegraphics[width=3in]{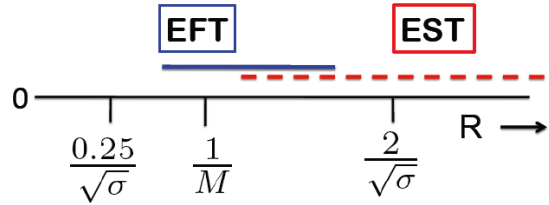}
\caption{Schematic showing  approximate domains of applicability of Effective Field Theory (EFT) (solid blue line) and Effective String Theory (EST) (red dashed line).
}
\label{cartoon}
\end{center}
\end{figure}

With use of analytic regularization \cite{Dietz}, string fluctuations do not renormalize the string tension $\sigma$, and  hence its physical interpretation as the energy per unit length of the classical flux tube is preserved. 
The leading  large distance correction to the heavy quark potential  is the L{\"u}scher term 
$ - \pi (d-2)/24 R\, $ \cite{Luscher}, which can be regarded as a renormalization of the intrinsic width at the intermediate distances shown as the region of overlap in Figure 2.

\section {Summary  and Future Work}

We have presented a physical picture of the $Z_N$ flux tubes giving rise to a confining string.  In this picture  net positive pressure in the interior of the $Z_N$ vortices  balances  net negative pressure outside.  (Perhaps at the deconfinement temperature the flux tube 'bursts'! ) We speculate that this general description is valid, independent of the details of the effective magnetic gauge theory from which it was obtained.

Comparison with lattice simulations at intermediate distances would test our hypothesis that there is an effective field theory underlying the confining string. 
Since $M \sim 3\, $ times $ T_C$, the SU(3) deconfinement temperature, the theory should be applicable for a range of temperatures  in the deconfined phase, where it
was used \cite{baker} in a preliminary study of spatial Wilson loops and where we expect some manifestation of the Higgs field.
\vspace{0.3cm}

{\bf{Acknowledgment:} }

 I would like to thank the organizers for the opportunity to participate in this very stimulating conference.

\end{document}